\documentclass[12pt,twoside]{article}
\usepackage{fleqn,espcrc1}

\newcommand{\be}{\begin{equation}}
\newcommand{\ee}{\end{equation}}
\newcommand{\ea}{\end{eqnarray}}
\newcommand{\baa}{\begin{eqnarray*}}
\newcommand{\eaa}{\end{eqnarray*}}
\newcommand{\bb}{\begin{thebibliography}}
\newcommand{\eb}{\end{thebibliography}}

\newcommand{\pup}{p}

\usepackage{graphicx}
\usepackage[figuresright]{rotating}

\newcommand{\AmS}{{\protect\the\textfont2pine
  A\kern-.1667em\lower.5ex\hbox{M}\kern-.125emS}}

\title{Investigation of single spin    
asymmetries in $\pi^{+}$ electroproduction\\ 
}

\author{K.A. Oganessyan\address{INFN-Laboratori Nazionali di Frascati, 
via Enrico Fermi 40, I-00044 Frascati, Italy}\address{DESY, Notkestrasse 
85, 22603 Hamburg, 
Germany}\thanks{e-mail: 
kogan@hermes.desy.de}\thanks{On leave of absence 
from Yerevan Physics Institute, Alikhanian 
Br.2, AM-375036 Yerevan, Armenia}, 
N.~Bianchi$^a$, E. De~Sanctis$^a$, 
W.-D.~Nowak\address{DESY Zeuthen, Platanenallee 6, D-15738 Zeuthen, Germany}
}
       
\begin{document}

\maketitle

\begin{abstract}
\small{
The azimuthal 
single 
target-spin asymmetries for $\pi^{+}$ production in semi-inclusive deep 
inelastic scattering of 
leptons off longitudinally polarized protons are evaluated using two main  
approaches available 
in the literature. It is shown that   
the approximation where the twist-2 {\it transverse} quark spin 
distribution in the {\it longitudinally} polarized nucleon is small 
enough to be neglected leads to a consistent description 
of all existing asymmetries observed by the HERMES experiment.   
}
\end{abstract}

\vspace*{0.5cm}

PACS: 13.85.Ni, 13.87.Fh, 13.88.+e

\section{Introduction}
\label{sec:intro} 
Semi-inclusive deep inelastic scattering (SIDIS) of leptons off a polarized 
nucleon  
\begin{equation}
l + \vec{p} \to l' + h + X
\label{R1}
\end{equation}
is an important process to study the 
internal structure of the nucleon and its spin properties. In 
particular, measurements of azimuthal distributions 
of the detected hadron provide valuable information on hadron structure 
functions, quark-gluon correlations and parton fragmentation functions. 

A significant target-spin asymmetry of the distributions in the azimuthal 
angle $\phi$ of the pion related to the lepton scattering plane for $\pi^{+}$ 
electroproduction in a {\it longitudinally} polarized hydrogen target has been 
recently observed by the HERMES collaboration~\cite{HERM,DELIA}. At the same 
time the SMC collaboration has studied the azimuthal distributions of pions 
produced in deep inelastic scattering off {\it transversely} polarized protons 
and deuterons~\cite{SMC}.  
These results have been  
interpreted as the effects of {\it naive} ``time-reversal-odd'' (T-odd) 
fragmentation 
functions \cite{COL,AK,TM,ARTRU,JAF1}, arising from non-perturbative 
hadronic final-state interactions. They have initiated a number of 
phenomenological 
approaches to evaluate these asymmetries using 
different input distribution and fragmentation functions 
\cite{BOER,BM0,EFR0,SNO,AD,KO,EP}. 
Actually, there are two main approaches in the literature which aim at 
explaining the experimental data:       

(i) The approximation where the twist-2 {\it transverse} quark spin 
distribution in the {\it longitudinally} polarized nucleon, 
$h_{1L}^{\perp(1)}(x)$, is considered small enough to be 
neglected~\cite{BOER,BM0,EFR0,SNO}. This results in good 
agreement with the Bjorken-$x$ behavior of the {\it sin}$\phi$ and 
{\it sin}$2\phi$ asymmetries observed at HERMES. 
Note, that this does not require the twist-3 interaction-dependent part 
of the fragmentation function, $\tilde{H}(z)$, to be zero. 

(ii) The approximation where the contribution of the interaction-dependent 
twist-3 term, $\tilde{h}_L(x)$, in the distribution function $h_L(x)$  
is assumed to be negligible, but  $\tilde{H}(z)$ 
is not constrained~\cite{BM0}. 

Another approximation, where only the twist-2 distribution 
and fragmentation functions are used, i.e. 
the interaction-dependent twist-3 parts of distribution and 
fragmentation functions are neglected, was proposed earlier~\cite{KO,EP}. 
For certain values of parameters this results in good agreement with the 
HERMES data~\cite{HERM}. However, it leads to the inconsistency that all 
T-odd fragmentation functions would be required to vanish~\cite{TM,ST}. 
Thus, in following we do not consider it anymore.  

In this paper we provide an analysis in the framework of the two above 
given approximations to 
evaluate the single target-spin {\it sin}$\,\phi$ and {\it sin}$\,2\phi$ 
asymmetries for $\pi^{+}$ production observed at HERMES~\cite{HERM,DELIA} in  
semi-inclusive deep inelastic scattering of leptons off longitudinally 
polarized protons. 

The paper is organized as follows: In Sec. \ref{sec: formul} we define 
the single spin azimuthal asymmetries and the involved twist-2 and twist-3 
distribution and fragmentation 
functions. In Sec. \ref{sec: res} we analyze the numerical results on the 
{\it sin}$\,\phi$ and {\it sin}$\,2\phi$ asymmetries. In 
Sec. \ref{sec: concl} we summarize the results.  
  
\section{Single spin asymmetries in $\gamma^{*} \vec{\pup} \to \pi X$}
\label{sec: formul}

\begin{figure}[ht]
\begin{center}
\includegraphics[width=10.cm]{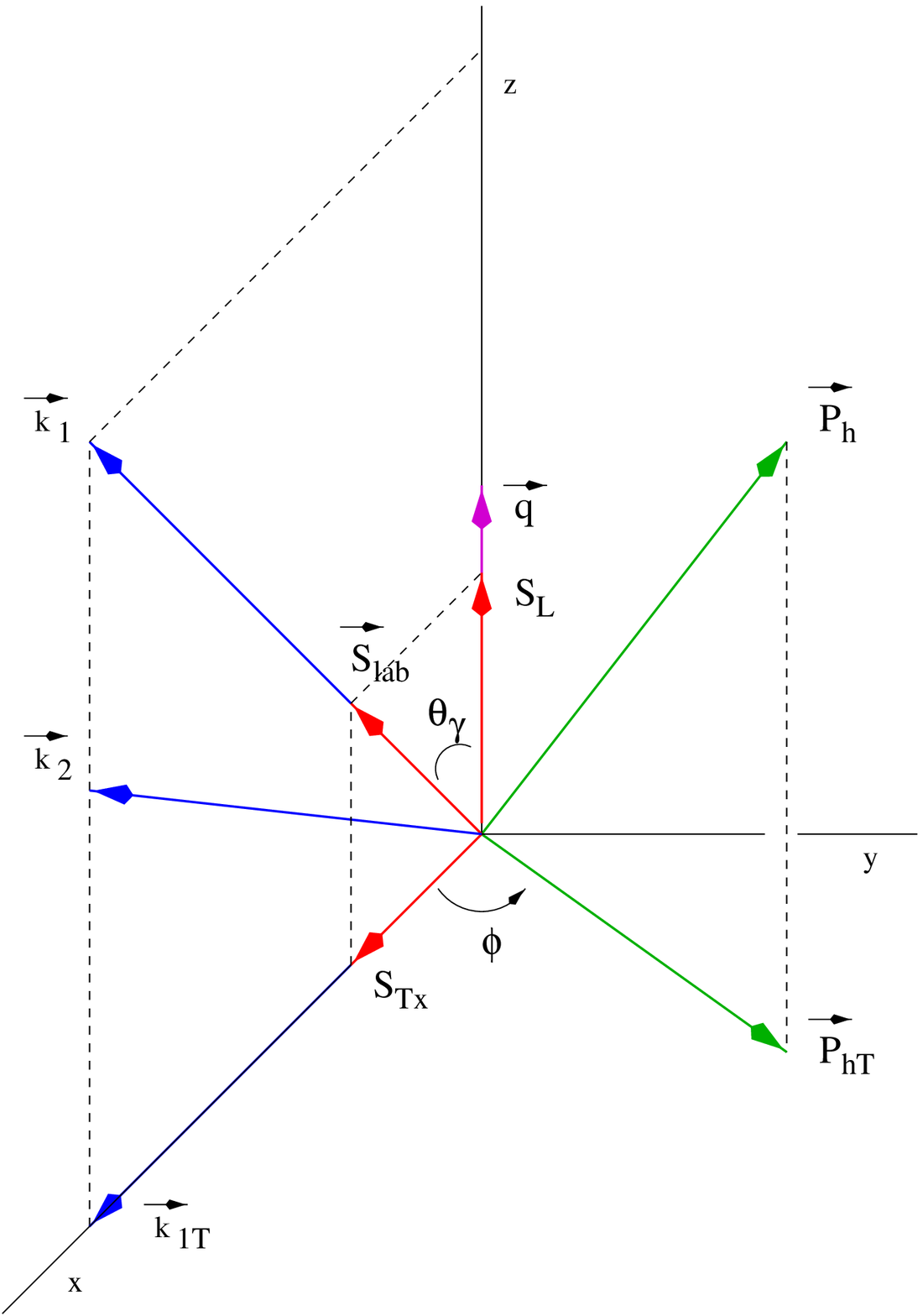}
\caption{
The kinematics of the process (\ref{R1}).
}
\label{az}
\end{center}
\end{figure}

The kinematics of the process (\ref{R1}) is illustrated in 
Fig.~\ref{az}: $k_1$ ($k_2$) is the 4-momentum of the 
incoming (outgoing) charged lepton, $Q^2=-q^2$, where $q=k_1-k_2$, 
is the 4-momentum of the virtual photon. 
$P$ ($P_h$) is the momentum of the target (observed) hadron, $x=Q^2/2(Pq)$, 
$y=(Pq)/(Pk_1)$, $z=(PP_h)/(Pq)$, $k_{1T}$ is the incoming lepton
transverse momentum with respect to the virtual photon momentum direction, 
and $\phi$ is the azimuthal angle between $P_{hT}$ and $k_{1T}$ around the 
virtual photon direction.  
Note that the azimuthal angle of the transverse 
(with respect to the virtual photon) component of the target polarization, 
$\phi_S$, is equal to 0 ($\pi$) for the target polarized parallel 
(anti-parallel) to the beam \cite{OABK}. 

The {\it sin}$\,\phi$ and {\it sin}$\,2\phi$ moments in the SIDIS 
cross-section can be related to the parton distribution and fragmentation 
functions involved in
the parton level description of the underlying process \cite{AK,TM}. 
These moments are defined as appropriately weighted integrals over 
$P_{hT}$ (the transverse momentum of the observed hadron) of 
the cross section asymmetry:
\begin{equation}
\langle \frac{{\vert P_{hT}\vert}}{M_h} \sin \phi \rangle \equiv 
\frac{\int d^2P_{hT} \frac{{\vert P_{hT}\vert}}{M_h}
\sin \phi \left(d\sigma^{+}-d\sigma^{-}\right)}
{\int d^2P_{hT} \left(d\sigma^{+} + d\sigma^{-}\right)},
\label{ASD1}
\end{equation}
\begin{equation}
\langle \frac{{\vert P_{hT}\vert}^2}{MM_h} \sin 2\phi \rangle \equiv 
\frac{\int d^2P_{hT} \frac{{\vert P_{hT}\vert}^2}{MM_h}
\sin 2\phi \left(d\sigma^{+}-d\sigma^{-}\right)}
{\int d^2P_{hT} \left(d\sigma^{+} + d\sigma^{-}\right)}.
\label{ASD2}
\end{equation}

Here $+ (-)$ denotes the anti-parallel (parallel) 
longitudinal polarization of the target and $M$ ($M_h$) is the mass of the 
target (final hadron).

The weighted single target-spin asymmetries defined above are related to 
the ones measured by HERMES~\cite{HERM} through the following relations:  
\begin{equation}
 A^{\sin\phi}_{UL} 
\approx {2M_h \over {\langle P_{hT}\rangle }} \langle \frac{\vert 
P_{hT}\vert}{M_h} \sin \phi_h \rangle, 
\label{REL1}
\end{equation}
\begin{equation}
A^{\sin2\phi}_{UL} 
\approx {2MM_h \over {\langle P^2_{hT}\rangle }} \langle \frac{\vert 
P^2_{hT}\vert}{MM_h} \sin 2\phi_h \rangle, 
\label{REL2}
\end{equation}
where the subscripts $U$ and $L$ indicate unpolarized beam and longitudinally 
polarized target, respectively.  

These asymmetries are given by~\cite{AK,TM,OABK}~\footnote{We omit the current 
quark mass dependent terms.}
\begin{equation}
\langle \frac{\vert P_{hT}\vert}{M_h} \sin \phi_h \rangle = 
{1 \over I_0} [I_{1L} + I_{1T}], 
\label{AS}
\end{equation}
\begin{equation}
\langle \frac{{\vert P_{hT}\vert}^2}{MM_h} \sin 2\phi_h \rangle = 
{8 \over I_0} S_L (1-y) \, h_{1L}^{\perp(1)}(x)
z^2 H_1^{\perp (1)}(z), 
\label{AS1}
\end{equation}
where 
\begin{equation}
I_0 = (1+(1+y)^2) f_1(x) D_1(z),
\end{equation}
\begin{eqnarray}
I_{1L} &=& 4S_L {M \over Q}\,(2-y)\sqrt{1-y} \, 
[ -2 h_{1L}^{\perp(1)}(x) z H_1^{\perp (1)}(z) \nonumber \\ 
&+& x \tilde{h}_L(x) z H_1^{\perp (1)}(z) - h^{\perp(1)}_{1L}(x) \tilde{H}(z) ],
\label{ASL}
\end{eqnarray}
\begin{equation}
I_{1T} = 2 S_{T\,x}\,(1-y)\,h_1(x) z H_1^{\perp (1)}(z). 
\label{AST}
\end{equation}

Here the components of the longitudinal and transverse target polarization 
in the virtual photon frame are denoted by $S_L$ and $S_{Tx}$, respectively.
Twist-2 distribution and fragmentation functions have a subscript `1':
$f_1(x)$ and $D_1(z)$ are the usual unpolarized distribution and 
fragmentation functions, while 
$h^{\perp (1)}_{1L}(x)$ and $h_1(x)$ describe the quark transverse spin 
distribution in longitudinally and transversely polarized nucleons,  
respectively. The interaction-dependent part of the twist-3 distribution 
function in the longitudinally polarized nucleon, $h_L(x)$~\cite{JJ91}, is 
denoted by $\tilde{h}_L(x)$~\cite{JJ92,MT}. 

For approach (i), where the twist-2 {\it transverse} quark spin 
distribution in the {\it longitudinally} polarized nucleon, 
$h_{1L}^{\perp(1)}(x)$, is set to zero, it follows:  
\be
h_L(x) = \tilde{h}_L(x) = h_1(x).
\label{HL4} 
\ee

In approach (ii) the interaction-dependent twist-3 part, 
$\tilde{h}_L(x)$, of the twist-3 distribution function, $h_L(x)$, is 
set to zero and hence~\cite{JJ92,MT}:
\be
h_{1L}^{\perp(1)}(x) = -x^2 \int_x^1 dy \frac{h_1(x)}{y^2}.
\label{HL3} 
\ee
It is worth noting here that 
calculations performed in the instanton model of Quantum 
Chromodynamics~\cite{DP} indicate the parametric smallness 
of the twist-3 contribution to the polarized structure function, 
while in the bag model this twist-3 contribution is comparable 
to the twist-2 contribution at small $Q^2$~\cite{JJ92,KK}.

The spin-dependent fragmentation function $H_1^{\perp (1)}(z)$, describing 
transversely polarized quark fragmentation~\cite{COL},  
correlates the transverse spin of a quark with a preferred transverse 
direction for the production of the pion.  
The fragmentation 
function $\tilde{H}(z)$ is the interaction-dependent part of the twist-3 
fragmentation function, $H(z)$,~\cite{TM} and is directly connected to 
$H_1^{\perp (1)}(z)$: 
\be
\tilde{H}(z)=z\frac{d}{dz}(zH_1^{\perp(1)}(z)). 
\ee
The distribution and fragmentation functions with superscript $(1)$ denote 
$p^2_T$-, $k^2_T$ - moments, where $p_T$, $k_T$ are the  
intrinsic transverse momenta of the initial and final quark, respectively.     

\section{Numerical results}
\label{sec: res}

For the numerical calculations in this work the non-relativistic approximation 
$h_1(x) = g_1(x)$ is used as a lower limit~\footnote{In the 
non-relativistic quark
model $h_1(x,\mu^2_0) = g_1(x,\mu^2_0)$. Several models suggest that  
$h_1(x)$ is close to $g_1$~\cite{JJ92,PP,BCD}. The evolution properties 
of $h_1$ and 
$g_1$, however, are very different~\cite{SV}. At the $Q^2$ values of the 
HERMES measurement 
the assumption $h_1=g_1$ is fulfilled at large, i.e. valence-like, $x$ values, 
while large differences occur at lower $x$~\cite{KNO}. }, 
and $h_1(x)=(f_1(x)+g_1(x))/2$ as an upper limit~\cite{SOF}.  
For the sake of simplicity, $Q^2$-independent parameterizations were chosen  
for the distribution functions $f_1(x)$ and $g_1(x)$~\cite{BBS}. 

To obtain the T-odd fragmentation function $H_1^{\perp (1)}(z)$,
the Collins ansatz~\cite{COL} for the analyzing power of 
transversely polarized quark fragmentation was adopted:
\begin{equation}
A_C(z,k_T) \equiv \frac{\vert k_T \vert}{M_h}\frac{H_1^{\perp}(z,k_T^2)}
{D_1(z,k_T^2)} = \eta \frac{M_C\,\vert k_T \vert}{M_C^2+k_T^2},
\label{H1T}
\end{equation}
where $\eta$ is taken as a constant, although, in principle it could be $z$ 
dependent.
 
For the distribution of the final parton intrinsic transverse momentum, 
$k_T$, in the unpolarized fragmentation function $D_1(z,k^2_T)$, a Gaussian 
parameterization was used~\cite{KM} with $\langle z^2 k_T^2 \rangle = b^2$ 
(in the numerical calculations $b = 0.36$ GeV was taken~\cite{PYTHIA}). 
For $D_1^{\pi^{+}} (z)$, the parameterization from Ref.~\cite{REYA} was 
adopted.
In Eq.(\ref{H1T}) $M_C$ is a typical hadronic mass whose value ranges from
$2m_{\pi}$ to $M_p$.  

It is important to point out here that the T-odd fragmentation function 
calculated with the Collins ansatz (Eq.(\ref{H1T})) at a  
reasonable value 
of the parameter $M_C$ ($M_C=2m_{\pi}$) with $\eta=1.0$, turns out to be 
in very good agreement with the parameterization~\cite{BM0,BL} that 
was based on a fit of $p p^{\uparrow} \to \pi X$ experimental 
data~\cite{ADAMS}. In our calculations a good agreement with 
HERMES results was achieved at that value of $M_C$, with $\eta=0.8$. 
This value of $\eta$ may indicate possible contributions of other   
mechanisms~\cite{ET,SQ} in $p p^{\uparrow} \to \pi X$.  

\begin{figure}[ht]
\begin{center}
\includegraphics[width=12.cm]{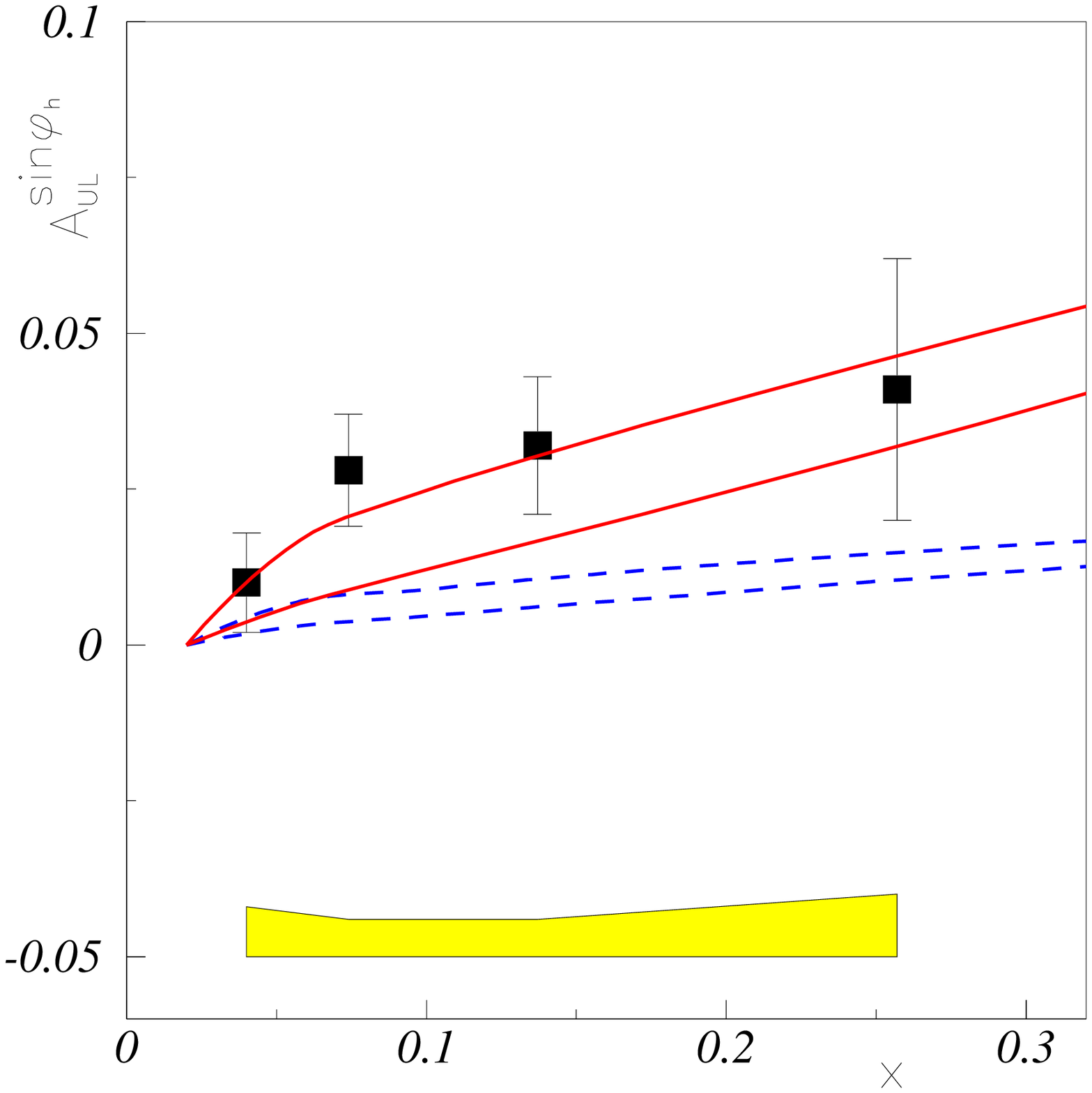}
\caption{
The single target-spin asymmetry $A^{\sin\phi}_{UL}$ for $\pi^{+}$ 
production as a function of Bjorken-$x$, evaluated using $M_C=2 m_{\pi}$  
and $\eta=0.8$ in Eq.(\ref{H1T}). The results obtained within 
approaches (i) and (ii) are denoted by pairs of full and dashed lines, 
respectively. 
For each approach two curves are presented corresponding to $h_1=g_1$ (lower 
curve) and 
$h_1=(f_1+g_1)/2$ (upper curve). HERMES data are from Ref.~\cite{HERM}.
}
\label{f1}
\end{center}
\end{figure}

In Fig.~\ref{f1}, the asymmetry $A^{\sin\phi}_{UL}(x)$ of 
Eq.(\ref{REL1}) for $\pi^{+}$ production on a proton target evaluated 
within the two approaches described in the Sec.~\ref{sec:intro},
is presented as a function of Bjorken-$x$ and compared to HERMES 
data~\cite{HERM}, which correspond to $1$ GeV$^2$ $\leq Q^2 \leq 15$ 
GeV$^2$, $4.5$ GeV $\leq E_{\pi} \leq 13.5$ GeV, $0.02 \leq x \leq 0.4$, 
$0.2 \leq z \leq 0.7$, and $0.2 \leq y \leq 0.8$. 

The curves have been calculated by integrating 
over the HERMES kinematic range taking $\langle P_{hT} \rangle 
= 0.365$ GeV and $\langle P_{hT}^2 \rangle 
= 0.165$ GeV$^2$ as input. The latter values are obtained in this
kinematic region assuming a Gaussian parameterization of 
the distribution and fragmentation functions 
with $\langle p_T^2 \rangle=(0.44)^2$ GeV$^2$~\cite{PYTHIA}. 

The results obtained within the approaches (i) and (ii) are denoted 
by pairs of full and dashed lines, respectively. 
For each approach two curves are presented according to the upper 
and lower limits chosen for $h_1(x)$. 

From Fig.~\ref{f1} it can be concluded that there is good agreement 
between results of the approach (i) and the HERMES data. The 
results of approach (ii) appear too low. 

Note that the `kinematic' contribution to $A^{\sin\phi}_{UL}(x)$, coming
from the transverse component of the target polarization with respect to
the virtual photon direction and given by $I_{1T}$ (Eq.(\ref{AST})),  
is about $(45-50) \%$ in approach (ii). In approach (i), 
where the twist-3 part of the fragmentation functions does not contribute 
(because of $h_{1L}^{\perp(1)}(x) \approx 0$)~\cite{BOER,BM0,EFR0,SNO}, it 
amounts to $25\%$. 

\begin{figure}[ht]
\begin{center}
\includegraphics[width=12.cm]{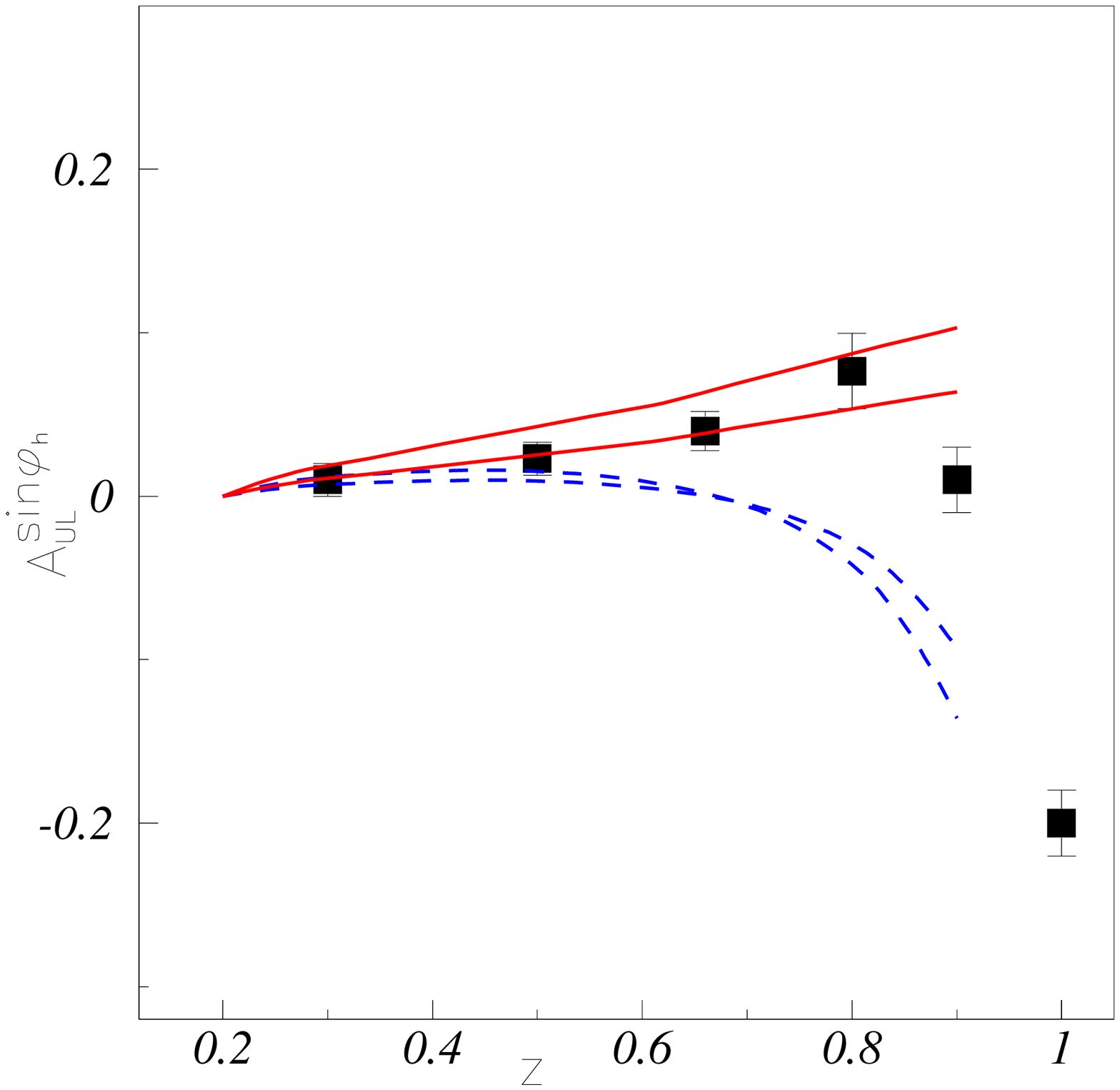}
\caption{
The single target-spin asymmetry $A^{\sin\phi}_{UL}$ for $\pi^{+}$ 
production as a function of $z$ evaluated using the same parameters as in 
Fig.~\ref{f1}. The curves have the same notations as in the 
Fig.~\ref{f1}. 
HERMES preliminary data not corrected for smearing (the error bars 
correspond to the statistical uncertainties only), are taken from 
Ref.~\cite{DELIA}.
}
\label{f2}
\end{center}
\end{figure}

The $z$ dependence of the asymmetry $A^{\sin\phi}_{UL}$ is shown in 
Fig.~\ref{f2} and compared with HERMES preliminary data~\cite{DELIA}, which 
here extend up to $z=1$.   
In approach (i) the {\it sin}$\,\phi$ asymmetry 
increases with $z$ 
and is in good agreement with data up to $z=0.8$. The sharp decrease  
of data for higher $z$ values reflects the transition from the 
semi-inclusive to the exclusive regime which requires a different 
investigation. In this respect our calculations are limited to $z \leq 0.9$. 
In approach (ii) the behavior of the asymmetry is quite 
different from that in approach (i) and fails to describe the experimental 
data already above $z \approx 0.5$. 
  
\begin{figure}[ht]
\begin{center}
\includegraphics[width=12.cm]{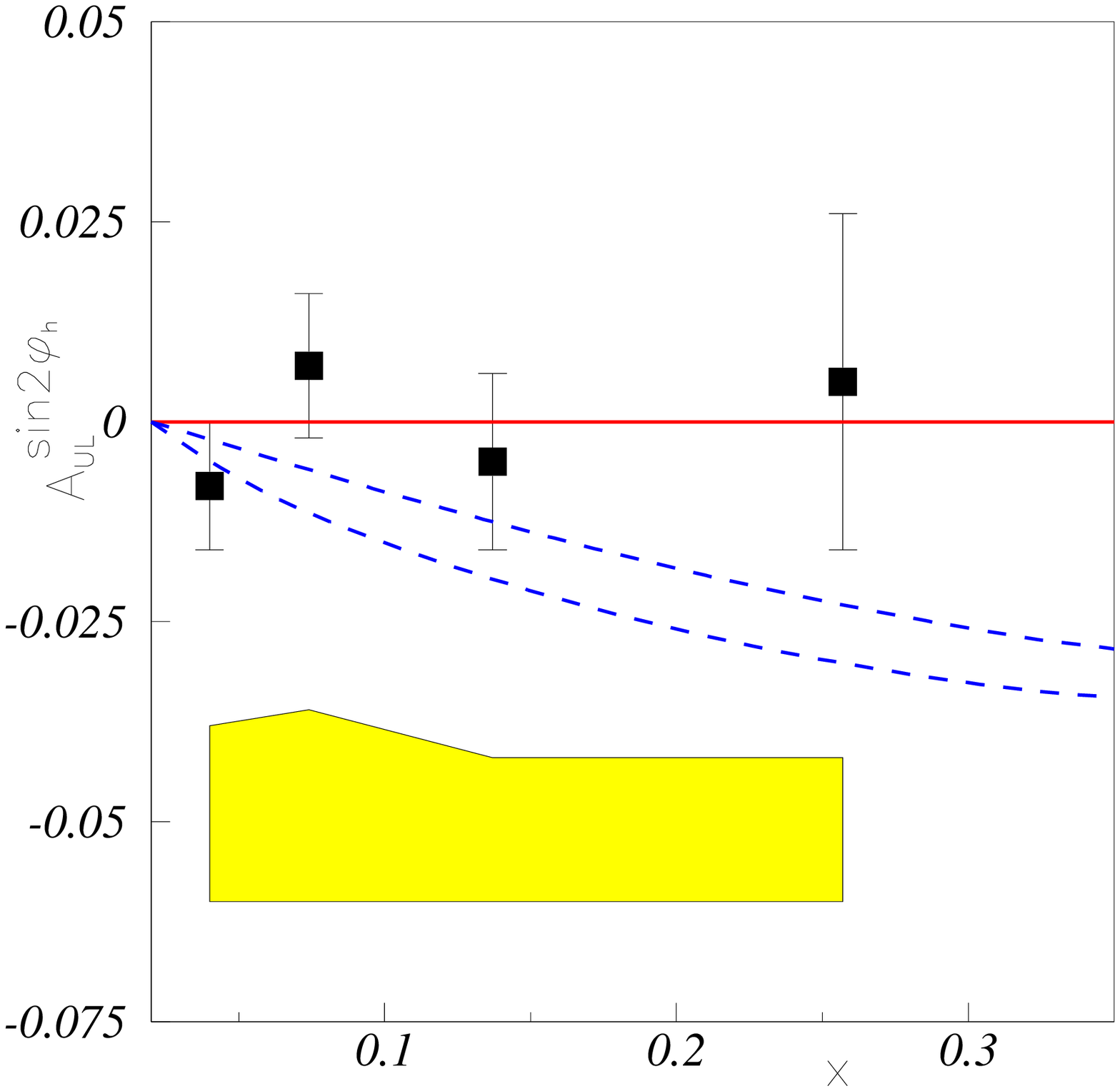}
\caption{
The single target-spin asymmetry $A^{\sin2\phi}_{UL}$ for $\pi^{+}$ 
production as a function of Bjorken $x$, evaluated using the same parameters 
as in Fig.~\ref{f1}. The curves have the same notations as in the 
Fig.~\ref{f1}; the line at $A^{\sin2\phi}_{UL} = 0$ corresponds to the result 
of approach (i). Data are from Ref.~\cite{HERM}.
}
\label{f3}
\end{center}
\end{figure}

Finally, in Fig.~\ref{f3} the asymmetry $A^{\sin2\phi}_{UL}$ of 
Eq.(\ref{REL2}) is presented for $\pi^{+}$ production on a proton target  
as a function of Bjorken-$x$ and compared to HERMES data~\cite{HERM}. 
Approach (ii), where $h_{1L}^{\perp(1)}(x) = 0$, 
leads directly to $A^{\sin2\phi}_{UL} = 0$ in better agreement with the data. 
The curves from approach (ii) are also compatible with the data, taking into 
account their total accuracy. Clearly, more accurate 
data will better constrain the phenomenological predictions.  
It is worth mentioning that by changing the parameters of 
the input functions it is possible to increase the magnitude of the  
$|A^{\sin\phi}_{UL}|$ asymmetry calculated in the approach (ii). However,  
this also modifies the $z$-dependence leading to a stronger disagreement 
with the data, and also increases the magnitude of the 
calculated asymmetry $|A^{\sin2\phi}_{UL}|$ deteriorating the compatibility 
with the data. 

In addition we note that approach (i) well describes   
the $P_{hT}$ dependence of $A^{\sin\phi}_{UL}$
observed at HERMES~\cite{QCD00}, too.  
 
\section{Conclusions} 
\label{sec: concl}

We have evaluated the $A^{\sin\phi}_{UL}$ and $A^{\sin2\phi}_{UL}$ 
single-spin asymmetries for semi-inclusive $\pi^{+}$ production in 
semi-inclusive deep 
inelastic scattering of leptons off longitudinally polarized protons 
using the two main approaches available in the literature. The 
results have been compared to the recent HERMES data. The 
approximation where the twist-2 {\it transverse} quark spin 
distribution in the {\it longitudinally} polarized nucleon, 
$h_{1L}^{\perp(1)}(x)$, is neglected, gives a consistent 
description of both $A^{\sin\phi}_{UL}(x)$ and $A^{\sin2\phi}_{UL}(x)$ 
and describes well the $z$-dependence of $A^{\sin\phi}_{UL}$. 

\section{Acknowledgments}
\label{sec: ackn}

We thank Daniel Boer, Bob Jaffe, Aram Kotzinian and Piet Mulders for 
interesting discussions. 
The work of K.A.O. was in part supported by INTAS contributions (contract 
numbers 93-1827 and 96-287) from the European Union.



\begin{thebibliography}{99}
  \bibitem{HERM} H.~Avakian, for the HERMES Collaboration, 
                 Nucl. Phys. (Proc. Suppl.) B{\bf 79} (1999) 
             523; HERMES Collaboration, A. Airapetian, et.al, Phys. 
             Rev. Lett. {\bf 84} (2000) 4047. 
  \bibitem{DELIA} D.~Hasch, for the HERMES collaboration, 35th Rencontres de 
                 Moriond: QCD and Hadronic Interactions, Les Arcs,
                 France, March 18 - 25, 2000. 
  \bibitem{SMC} A.~Bravar, Nucl. Phys. (Proc. Suppl.) B{\bf 79} (1999) 520. 
  \bibitem{COL} J.~Collins, Nucl. Phys. B{\bf 396} (1993) 161.
  \bibitem{AK} A.~Kotzinian, Nucl. Phys. B{\bf 441} (1995) 234.
  \bibitem{TM} P.J.~Mulders and R.D.~Tangerman, Nucl Phys. B{\bf 461} 
(1996) 197. 
  \bibitem{ARTRU} X.~Artru, J.~Czyzewski and H.~Yabuki, Z. Phys C{\bf 73}, 
               (1997) 527. 
  \bibitem{JAF1} R.~Jaffe, X.~Ji and J.~Tang, Phys. Rev. Lett. {\bf 80} (1998) 
                1166. 
  \bibitem{BOER} D.~Boer, hep-ph/9912311. 
   \bibitem{BM0}  M. Boglione and P.J. Mulders, Phys. Lett. B{\bf 478} 
(2000) 114.  
  \bibitem{EFR0} A.V.~Efremov, hep-ph/0001214. 
  \bibitem{SNO}  E.~De~Sanctis, W.-D.~Nowak, and K.A.~Oganessyan, 
                 Phys. Lett. B{\bf 483} (2000) 69. 
  \bibitem{AD} M.~Anselmino and F.~Murgia, hep-ph/0002120.
  \bibitem{KO} A.M.~Kotzinian, K.A.~Oganessyan, A.R.~Avakian, and 
               E.~De Sanctis, 
               Nucl. Phys. A{\bf 666-667} (2000) 290.  
  \bibitem{EP} A.V.~Efremov, M.~Polyakov, K.~Goeke, and D.~Urbano, 
               Phys. Lett. B{\bf 478} (2000) 94. 
  \bibitem{ST} A.~Schaefer and O.V.~Teryaev, Phys. Rev. D{\bf 61} (2000) 
               077903. 
  \bibitem{OABK} K.A. Oganessyan, A.R. Avakian, N. Bianchi, and 
                 A.M. Kotzinian, 
                hep-ph/9808368;  Proceedings of the workshop Baryons'98, 
                Bonn, Sept. 22-26, 1998.
  \bibitem{JJ91} R.~Jaffe, and X.~Ji, Phys. Rev. Lett, {\bf 67} (1991) 552.
  \bibitem{JJ92} R.~Jaffe, and X.~Ji, Nucl. Phys. B{\bf 375} (1992) 527.
  \bibitem{MT} R.D.~Tangerman, and P.J.~Mulders, NIKHEF-94-P7, hep-ph/9408305. 
  \bibitem{DP} B.~Dressler, and M.V.~Polyakov, Phys.Rev. D{\bf 61} 
               (2000) 097501. 
  \bibitem{KK} Y.~Kanazawa, and Y.~Koike, Phys.Lett. B{\bf 403} (1997) 357.
  \bibitem{PP} P.V.~Pobylitsa and M.V.~Polyakov, Phys. Lett. {\bf B389} 
           (1996) 350.  
  \bibitem{BCD} V.~Barone, T.~Calarco and A.~Drago, Phys. Lett. B{\bf 390} 
           (1997) 287.
  \bibitem{SV} S.~Scopetta, and V.~Vento, Phys.Lett. B{\bf 424} (1998) 25.
  \bibitem{KNO}  V.A.~Korotkov, W.-D.~Nowak, and K.A.~Oganessyan, 
                 hep-ph/0002268; DESY 99-176.
  \bibitem{SOF} J.~Soffer, Phys. Rev. Lett. {\bf 74} (1995) 1292.
  \bibitem{BBS} S.~Brodsky, M.~Burkardt, and I.~Schmidt, 
                 Nucl. Phys. B{\bf 441} (1995) 197. 
  \bibitem{KM} A.M.~Kotzinian, and P.J.~Mulders, Phys. Lett. B{\bf 406} (1997) 373.  
  \bibitem{PYTHIA} T.~Sjostrand, Comp. Phys. Commun. {\bf 82} (1994) 74; 
              CERN-TH.7112/93; hep-ph/9508391. 
  \bibitem{REYA} E.~Reya, Phys. Rep. {\bf 69} (1981) 195. 
  \bibitem{BL} M.~Boglione, and E.~Leader,  hep-ph/9911207.
         (1999) 054007.
  \bibitem{ADAMS} D.L.~Adams, et.al., Phys. Lett. B{\bf 261} (1991) 201; 
               Phys. Lett. B{\bf 264} (1991) 462.
  \bibitem{ET} A.V.~Efremov, O.V.~Teryaev, Phys. Lett. B{\bf 150} (1985) 
               383.  
  \bibitem{SQ} J.~Qiu, G.~Sterman, Phys. Rev. Lett. {\bf 67} (1991) 2264; 
                Nucl. Phys. B{\bf 378} (1992) 52.  
  \bibitem{QCD00}  K.A.~Oganessyan, N.~Bianchi, E.~De~Sanctis, 
                   and W.-D.~Nowak, hep-ph/0010063; Proceedings of the 
                   Euroconference QCD'00, 6-13 July, 2000, Montpellier. 

  \end{thebibliography}
\end{document}